\documentclass[12pt]{article}
\usepackage{amsfonts,latexsym,fullpage,amsmath,latexsym,amssymb,epsf}

\newcommand{\ket}[1]{|#1\rangle} \newcommand{\bra}[1]{\langle #1|}

\newcommand{\beq}{\begin{equation}} \newcommand{\eeq}{\end{equation}}
\newcommand{\beqy}{\begin{eqnarray}}
\newcommand{\eeqy}{\end{eqnarray}}

   \newenvironment{Definition*}{{\bf
Definition}}{} \def\C{{\mathbb{C}}} 
  
 \newcommand{\cH}{{\cal H}}
 
\newcommand{\cP}{{\cal P}}

\begin{document}

\date{August 19, 2003}
\title{Quantum circuits for single-qubit 
measurements corresponding to platonic solids}

\author{Thomas Decker\thanks{e-mail: {\protect \{decker,
janzing\}@ira.uka.de}}, Dominik Janzing, and Thomas Beth \\ \small
Institut f{\"u}r Algorithmen und Kognitive Systeme, Universit{\"a}t
Karlsruhe,\\[-1ex] \small Am Fasanengarten 5, D-76\,131 Karlsruhe,
Germany}

\maketitle

\abstract{Each platonic solid defines a single-qubit positive operator
valued measure (POVM) by interpreting its vertices as points on the
Bloch sphere. We construct simple circuits for implementing this kind
of measurements and other simple types of symmetric POVMs on one
qubit. Each implementation consists of a discrete Fourier transform
and some elementary quantum operations followed by an orthogonal
measurement in the computational basis.}

\section{Introduction}
A key postulate of textbook quantum mechanics is the assumption that
measurements correspond to self-adjoint operators $A$ in such a way
that the probability of each possible measurement outcome or set of
possible outcomes can be computed from the spectral projections of
$A$. If the corresponding system Hilbert space is finite dimensional
$A$ can be written as $A=\sum_j \lambda_j P_j$ where $P_j$ is 
the projection
on the eigenspace with eigenvalue $\lambda_j$.  The probability
of the outcome $\lambda_j$ is ${\rm tr}(\rho P_j)$ if the system is in a state
with density matrix $\rho$. This type of measurement is called 
{\it von-Neumann measurement}, 
{\it orthogonal measurement} or {\it projector-valued measurement}.

Within the standard model of a quantum computer one can easily show
that it is in principle possible to implement measurements for all
self-adjoint operators $A$ acting on the Hilbert space
$(\C^2)^{\otimes l}$, i.e., the state space of a quantum register
with $l$ qubits.  Since a universal quantum computer allows the
implementation of each unitary transformation one could perform a
unitary operation $U$ that diagonalizes $A$ with respect to the
computational basis and measure with respect to this basis.

However, the description of measurements by self-adjoint operators is
not general enough. Most general measurements are described by
positive operator valued measures (POVMs). A
POVM is defined as follows \cite{Davies2}. Let $\Omega$ be the set of possible
outcomes and $\Sigma$ be a sigma-algebra of measurable subsets of
$\Omega$. Let $\cP$ be the set of positive operators acting on the
Hilbert space $\cH$. Then a POVM $A$ is a map $A: \Sigma \rightarrow
\cP, m \mapsto A_m,$ with the following properties:
\begin{enumerate}
\item For all countable families $(m_j)$ of mutually disjoint sets $m_j$
one has 
\[
A_{\cup_j m_j} = \sum_j A_{m_j}\,,
\]
where the infinite sum converges in the weak operator topology.
\item
$A_{\Omega}={\bf 1}$. 
\end{enumerate}
The probability for obtaining an outcome in the set $m$ is given by
${\rm tr}(\rho A_m)$. When the set $\Omega$ of possible outcomes is
finite or countably infinite a POVM is uniquely given by a family
$(A_j)$ of positive operators such that $p_j={\rm tr}(\rho A_j)$ is
the probability for obtaining the outcome $j$.
We only consider POVMs with a finite set $\Omega$ of
outcomes. Furthermore, the considered POVMs have the following properties:

\begin{enumerate}

\item The family $(A_j)$ describes a single-qubit measurement, i.e.,
the system Hilbert space is $\C^2$.

\item Each $A_j$ is a rank-one operator, i.e.,
$A_j=\ket{\Psi_j}\bra{\Psi_j}$. The vectors $\ket{\Psi_j}$ have the
same length. They are not necessarily normalized.


\item The operators $A_j$ correspond to symmetric points on the
Bloch sphere. The symmetry groups are finite subgroups of $SO(3)$. 
The possible symmetry groups are the cyclic and dihedral groups and
the symmetry groups of the platonic solids. 

\end{enumerate}

These properties show that we restrict our attention to a rather
specific class of symmetric POVMs. The symmetry is fundamental in our
constructions of the circuits implementing the POVMs.  Specifically,
we choose a cyclic subgroup of the symmetry group corresponding to a
POVM. Under the action of the cyclic group the set of points on the
Bloch sphere decomposes into several orbits.  As shown in Section
\ref{Section 2.1} POVMs given by a single orbit can easily be
implemented by a discrete Fourier transform.  Since we have several
orbits we have to use additional gates besides the Fourier transform
to implement the POVM. This explains why the discrete Fourier
transform plays a central role in all constructed circuits.

The intention of this paper is to show how the symmetry of a POVM can
be used to construct a simple circuit for implementing the POVM.
To our knowledge,
there are no considerations of the implementation of POVMs besides
\cite{Sasaki}.  The investigation of the implementation 
and its complexity is motivated by the fact that there are examples
where generalized measurements can extract more information about an
unknown quantum state than projector-valued measurements.  Symmetric
POVMs may, for instance, be interesting when we want to distinguish
between symmetric states \cite{Sasaki}.  Furthermore, POVMs may
perform better than orthogonal measurements with respect to
appropriate information criteria (e.g. mutual information
\cite{Davies} or the least square error \cite{Eldar}).  Here we do
neither consider these ''quality'' criteria nor the post-measurement
state. The post-measurement state may be relevant in order to
understand information-disturbance trade-off relations
\cite{Fuchs}. 

In the next section we describe the basic principles for implementing
arbitrary POVMs. In Section \ref{Section 2} we specify the
correspondence of POVM operators to points on the Bloch
sphere. Furthermore, we specify the symmetry of POVMs. In Sections
\ref{Section 2.1} and \ref{Section 2.2} we consider the implementation
of POVMs with a cyclic or dihedral symmetry group, respectively. These
considerations are the basis of the implementations of POVMs
corresponding to platonic solids. The implementation of these
POVMs is discussed in Sections \ref{S 4}--\ref{S ico}.

\section{Orthogonal measurement of POVMs}
\label{Sektion ortho}
In this section we briefly rephrase Neumark's theorem
describing the reduction of POVMs to orthogonal measurements
\cite{Peres}.  This theorem allows to implement POVMs by performing
unitary transformations on the joint system consisting of the system to
be measured and an ancilla register. The unitary transformations are
followed by an orthogonal measurement in the computational basis.
 
Let $(A_j)$ with $j \in \{1, \ldots , n \}$ be a POVM with
corresponding Hilbert space ${\mathbb C}^d$ where each
$A_j=\ket{\Psi_j} \bra{\Psi_j} \in {\mathbb C}^{d \times d}$ is a
positive operator of rank one.  Due to the properties of POVMs we have
$\sum_j A_j =I_d$ where $I_d$ denotes the identity matrix of size $d$.
The choice of corresponding vectors $\ket{\Psi_j}$ is not unique since
we can multiply each $\ket{\Psi_j}$ with a phase factor that is
physically irrelevant. It is therefore reasonable to choose the
phase factors in such a way that the implementation of the POVM is
simplified. Our constructions in Sections~4--10 implicitly make use of
this.  For $n > d$ the vectors $\ket{\Psi_j}$ cannot be mutually
orthogonal. As a simple example we consider a system with Hilbert space
${\mathbb C}^2$ and the following vectors:
$$\ket{\Psi_1} = \sqrt{\frac{1}{3}}
\left(\begin{array}{c}1\\1\end{array}\right), \;\; \ket{\Psi_2} =
\sqrt{\frac{1}{3}}\left(\begin{array}{c}1\\ \omega
\end{array}\right) \;{\rm and} \;\; \ket{\Psi_3}=\sqrt{\frac{1}{3}}
\left(\begin{array}{c}1 \\ \omega^2\end{array} \right).$$ 
Here is $\omega:= {\rm exp}(-2\pi i /3)$ a third root of unity. We 
therefore have $$A_1 = \frac{1}{3}\left( \begin{array}{cc}1 & 1 \\ 1 & 1 
\end{array} \right), \; \; A_2=\frac{1}{3} \left( \begin{array}{cc}1 & 
\omega^2 \\ \omega & 1 \end{array} \right)\; {\rm and} \;\; A_3= 
\frac{1}{3}\left( 
\begin{array}{cc}1 & \omega \\ \omega^2 & 1 \end{array} \right)$$ as POVM 
operators. In Section \ref{Section 2.1} we consider a generalization
of this POVM.

Assuming orthogonal measurements as basic measurements, we have to
extend the system by at least $n-d$ dimensions to make a measurement
with $n$ different measurement outcomes possible. In order to simplify
notation, we consider the given system with $d$ dimensions as a
subsystem of a system with $n$ dimensions. Since we are interested in
quantum circuits we have to embed the system into a qubit
register. This can be done by assuming that the POVM consists of
$n=2^l$ operators. Note that this is no loss of generality since we
can extend a given POVM by an appropriate number of zero operators
$A_j = 0_d \in {\mathbb C}^{d \times d}$ where $0_d$ denotes the zero
matrix of size $d$.  This extension does not change the probability
distribution of the POVM since $p_j = {\rm tr}(\rho 0_d) = 0$ for a
zero operator $A_j=0_d$. In our example above we add the zero operator
$A_4 = 0_2$ to the three POVM operators. We obtain a POVM that can be
implemented on a register of two qubits.

The basic idea of Neumark's theorem is to implement an orthogonal
measurement $({\tilde A}_j)$ on the extended system with $n$
dimensions that corresponds to the POVM $(A_j)$ in the sense that it
reproduces the correct probabilities $p_j$.  We now consider the
construction of the orthogonal measurement $({\tilde A_j})$. Let $\rho
\in {\mathbb C}^{d \times d}$ be the density matrix of the state to be
measured. Then the state of the extended system with $n$ dimensions
can be written as ${\tilde \rho} = \rho \oplus 0_{n-d} \in {\mathbb
C}^{n \times n}$.  When we write the vectors $\ket{\Psi_j}$ as columns
of the matrix $$M= \left( \ket{\Psi_1} \ldots \ket{\Psi_n} \right) \in
{\mathbb C}^{d \times n},$$ the operators ${\tilde A}_j = \ket{{\tilde
\Psi}_j} \bra{{\tilde \Psi}_j} \in {\mathbb C}^{n \times n}$ are given
by $\ket{{\tilde \Psi}_j} = \ket{\Psi_j} \oplus \ket{\Phi_j} \in
{\mathbb C}^n$. The extended vectors $\ket{{\tilde \Psi}_j}$ are the
columns of the matrix
$${\tilde M} = \left( \begin{array}{ccc} \ket{\Psi_1} & \ldots &
\ket{\Psi_n} \\ \ket{\Phi_1} & \ldots & \ket{\Phi_n} \end{array}
\right) \in {\mathbb C}^{n \times n},$$ that is an arbitrary unitary
matrix containing $M$ as upper part of size $d \times n$.  The
extension of $M$ to a unitary matrix ${\tilde M}$ is always possible since 
the rows of $M$ are orthonormal. This is guaranteed by the fact that
each POVM $(A_j)$ satisfies $\sum_j A_j =I_d$. The probability distribution
${\tilde p}_j = {\rm tr}({\tilde \rho} {\tilde A}_j)$ equals the
distribution $p_j$ of the original POVM since $${\tilde p}_j = {\rm
tr}\left({\tilde \rho} {\tilde A}_j \right) = {\rm tr}\left(
\left(\rho \oplus 0_{n-d} \right)  \left(\begin{array}{cc}
\ket{\Psi_j}\bra{\Psi_j} & \ket{\Psi_j} \bra{\Phi_j} \\ \ket{\Phi_j}
\bra{\Psi_j} & \ket{\Phi_j} \bra{\Phi_j}
\end{array} \right) \right) = {\rm tr}\left( \rho A_j \right) = p_j.$$
In our example, we have $$M= \sqrt{\frac{1}{3}} \left( \begin{array}{cccc} 1 &
1 & 1 & 0 \\ 1 & \omega & \omega^2 & 0 \end{array} \right) \in
{\mathbb C}^{2 \times 4}.$$ A possible unitary extension ${\tilde M}$
of this matrix is given by $${\tilde M} = \sqrt{\frac{1}{3}} \left(
\begin{array}{cccc} 1 & 1 & 1 & 0 \\ 1 & \omega & \omega^2 & 0 \\ 1 &
\omega^2 & \omega & 0 \\ 0 & 0 & 0 & \sqrt{3} \end{array} \right) \in
{\mathbb C}^{4 \times 4}$$ leading to the vectors $\ket{ {\tilde
\Psi}_1} = \sqrt{1/3}(1,1,1,0)^T$, $\ket{ {\tilde \Psi}_2} =
\sqrt{1/3}(1,\omega ,\omega^2,0)^T$, $\ket{ {\tilde \Psi}_3} =
\sqrt{1/3}(1,\omega^2,\omega,0)^T$, and $\ket{ {\tilde \Psi}_4} =
(0,0,0,1)^T$.  With the state ${\tilde \rho} = \rho \oplus
0_2 \in {\mathbb C}^{4 \times 4}$, for instance,  we obtain the probability
$${\tilde p}_2 = {\rm tr}\left( \left( \begin{array}{cc|cc} \rho_{11}
& \rho_{12} & 0 & 0 \\ \rho_{21} & \rho_{22} & 0 & 0 \\ \hline0 & 0 &
0 & 0 \\ 0 & 0 & 0& 0 \end{array} \right)  \left(
\begin{array}{cccc} 1 & \omega^2 & \omega & 0 \\ \omega & 1 & \omega^2
& 0 \\ \omega^2 & \omega & 1 & 0 \\ 0 & 0 & 0 & 0 \end{array} \right)
\right) = {\rm tr}\left( \rho \left( \begin{array}{cc} 1 &
\omega^2 \\ \omega & 1 \end{array} \right) \right) = p_2$$ for the
second POVM operator. The probabilities for the other POVM
operators are computed similarly.

The implementation of a POVM with corresponding matrix $M$ is obtained
by the orthogonal measurement in the computational basis after
performing the unitary transformation ${\tilde M}^\dagger$ on the
system with initial state ${\tilde \rho}$. This unitary operation
maps the vector $\ket{ {\tilde \Psi}_j} = \ket{\Psi_j} \oplus
\ket{\Phi_j}$ to the computational basis vector $\ket{j}$.  Therefore,
the measurement in the computational basis after applying ${\tilde M}^\dagger$
corresponds to the measurement in the basis defined by the vectors
$\ket{{\tilde \Psi}_j}$.

In summary, we are interested in constructing and implementing the
matrix ${\tilde M}^\dagger$ for a given POVM corresponding to the
matrix $M$. In the following sections the construction of the matrices
${\tilde M}^\dagger$ is considered for symmetric POVMs besides the
decomposition of ${\tilde M}^\dagger$ into elementary (one- and
two-qubit) gates. The symmetry leads to simple constructions and
implementations based on Fourier transforms.

\section{Symmetric POVMs on a single qubit}
\label{Section 2}
As described in the previous section, the basis of the orthogonal
measurement of a POVM with corresponding matrix $M$ is the
implementation of ${\tilde M}^\dagger$. ${\tilde M}$ is
a unitary extension of $M$. We can apply the algorithm in Section
4.5.1 of \cite{Nielsen} to obtain a quantum circuit for ${\tilde
M}^\dagger$.  The algorithm decomposes the matrix ${\tilde M}^\dagger$
into a product of two-level matrices that can be translated into a
sequence of elementary gates, i.e., each gate operates on one or two
qubits. In general, the constructed circuit for ${\tilde M}^\dagger$
is of exponential size in the number $n$ of POVM operators.
Intuitively, some symmetry properties of the considered POVMs may
lead to algorithms constructing smaller circuits than the standard
algorithm that works for arbitrary unitary matrices. 

To specify the symmetry of POVMs on a single qubit using geometric
concepts, we use the correspondence of POVMs to points on the
Bloch sphere as already mentioned in the introduction.  Usually, each
point on the Bloch sphere is considered as a pure state.
Specifically, a pure state $\rho \in{\mathbb C}^{2\times 2}$
corresponds to the point $(x,y,z)^T \in {\mathbb R}^3$ on the Bloch
sphere with $$\left(
\begin{array}{c}x\\y\\z \end{array} \right) = \left( 
\begin{array}{c} {\rm tr}(\sigma_x \rho ) \\ {\rm tr}
(\sigma_y \rho) \\ {\rm tr}( \sigma_z \rho) \end{array} \right)$$
where $$\sigma_x=\left( \begin{array}{cc}0&1\\1&0\end{array}\right),
\sigma_y=\left(\begin{array}{cc}0&-i\\i&0\end{array}\right), \; {\rm
and} \; \sigma_z=\left( \begin{array}{cc}1&0\\0&-1\end{array}
\right)$$ denote the Pauli spin matrices. Conversely, the point
$(x,y,z)^T \in {\mathbb R}^3$ on the Bloch sphere corresponds to the
density matrix
$$\frac{1}{2}\left(\begin{array}{cc}1+z & x-iy \\ x+iy & 1-z
\end{array} \right) \in {\mathbb C}^{2\times 2}.$$
For some special states the points on the Bloch sphere are shown in
Figure
\begin{figure}
  \centerline{\epsffile{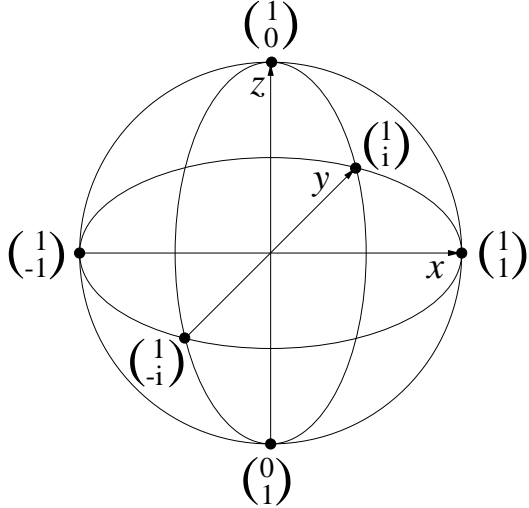}}
  \caption{Points on the Bloch sphere for some (unnormalized) state vectors.}
  \label{Figur Blochkugel}
\end{figure}
\ref{Figur Blochkugel}.  We now extend the Bloch sphere representation
for states to a representation of POVM operators of rank one. Note
that each pure state is a projection $\rho=\ket{\Psi}\bra{\Psi}$ of
rank one. By rescaling a POVM operator $A_j =
\ket{\Psi_j}\bra{\Psi_j}$ to a density matrix we can identify $A_j$
with a point on the Bloch sphere using the correspondence for pure states.

A symmetry of a POVM can be defined by a symmetry of the corresponding
points on the Bloch sphere. We are interested in POVMs with a finite
symmetry group for the points on the Bloch sphere, i.e., we consider
finite subgroups of $SO(3)$ \cite{Sternberg}. There are two infinite
families of finite subgroups, namely the cyclic groups $C_m$ and the
dihedral groups $D_m$ for $m\geq 2$ as symmetry groups of an $m$-sided
regular polygon (with the special case $m=2$). Furthermore, we have
the symmetry groups of the five platonic solids (tetrahedron, cube,
octa-, dodeca-, and icosahedron). As a restriction
for the latter symmetry groups, we assume the points on the Bloch
sphere of a POVM to coincide with the vertices of the platonic solid
corresponding to the symmetry group.

The vertices of the regular polygons and the platonic solids depend on
the orientation of the polygons and platonic solids in the Bloch
sphere.  Each orientation leads to another POVM. In order to simplify the
constructions in the following sections we choose specific
orientations of the regular polygons and platonic solids. To obtain
the implementation of a POVM corresponding to the same polygon or
platonic solid with another orientation, it suffices to implement a
single qubit operation on the qubit to be measured. 


\section{Cyclic groups}
\label{Section 2.1}
The simplest finite symmetry groups of points on the Bloch sphere are
the cyclic groups. For a fixed $m\geq 2$ we consider the rotations of
an $m$-sided regular polygon with a common axis perpendicular to the
face of the polygon. 
The rotations form a group that is isomorphic to the group $C_m=
\langle r \rangle$ with $r^m=1$.  The implementation of POVMs
corresponding to a single orbit of points under 
the action of a cyclic symmetry group is the basis of all
constructions in the following sections.

\begin{figure}
  \centerline{\epsffile{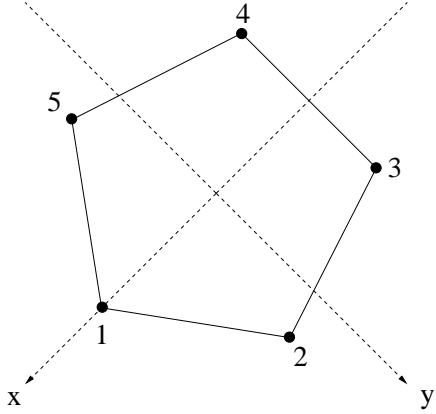}}
  \caption{Points of the cyclic POVM in the $xy$-plane for $m=5$.}
  \label{Figur cyclic vectors}
\end{figure}

In principle, we can choose an arbitrary orientation of the polygon
corresponding to the cyclic symmetry group. For simplification, we
choose the face of the regular polygon to be perpendicular to the
$z$-axis in the Bloch sphere. In other words, the $z$-axis is the
common axis of the rotations. For instance, the $5$-sided regular
polygon is shown in Figure \ref{Figur cyclic vectors}.

The cyclic symmetry group of the polygon is generated by the $2\pi/m$
rotation about the $z$-axis. In the Hilbert space, this rotation of
the Bloch sphere corresponds to the matrix ${\rm diag}(1,\omega) \in
{\mathbb C}^{2 \times 2}$ where $\omega := {\rm exp}(- 2 \pi i/m)$ is
an $m$th complex root of unity. The diagonal form of the matrix is the
reason for choosing the $z$-axis as common rotation axis.  If we
choose the vector $(1,1)^T \in {\mathbb C}^2$ and consider the orbit
under the symmetry group then we get the vectors $(1, \omega^{j})^T
\in {\mathbb C}^2$ for $j \in \{0, \ldots, m-1\}$. Other vectors of
the Bloch sphere do not lead to POVMs or to POVMs that correspond to a
polygon with another rotation. The latter case is discussed at the end
of the previous section. Due to the identity
$$\sum_{j=0}^{m-1}  \left(
\begin{array}{c} 1 \\ \omega^j \end{array} \right)
(1, \omega^{-j}) = \sum_{j=0}^{m-1} \left(
\begin{array}{cc}1&\omega^{-j}\\ \omega^{j}&1 \end{array} \right) = m
 I_2$$ the elements $\ket{\Psi_j}= \sqrt{1/m}(1,\omega^{j-1} )^T \in
{\mathbb C}^2$ for $j \in \{1, \ldots, m\}$ define a POVM with $m=n$
operators on a qubit.  Therefore, the unitary matrix $\tilde{M} \in
{\mathbb C}^{m \times m}$ is a unitary extension of the matrix
$$M=\sqrt{\frac{1}{m}} \left( \begin{array}{cccc}
1&1&\ldots&1\\1&\omega& \ldots & \omega^{m-1} \end{array} \right) \in
{\mathbb C}^{2 \times m}.$$ $M$ corresponds to the first two rows of
the discrete Fourier matrix
$$F_m=\sqrt{\frac{1}{m}}\left(\omega^{jk}\right)_{j,k=0}^{m-1}
\in {\mathbb C}^{m \times m}$$ of size $m$. Consequently, by considering
the qubit to be measured as a subsystem of an $m$-dimensional system 
the implementation of the inverse Fourier transform 
${\tilde M}^\dagger = F_m^\dagger$ leads to the probability
distribution of the cyclic POVM on the qubit.

\begin{figure}
  \centerline{\epsffile{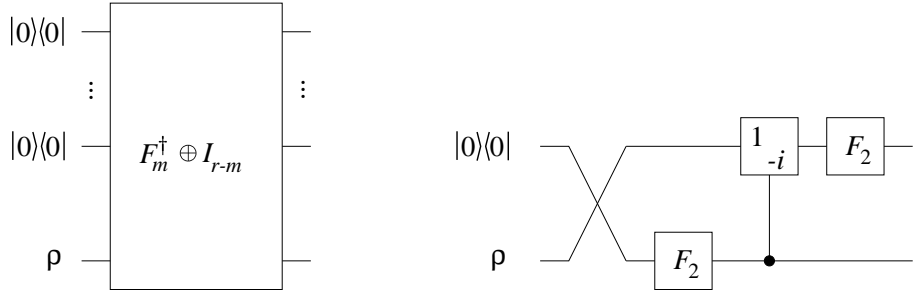}} \caption{The general circuit
  scheme for implementing the cyclic POVM (left side) and the circuit
  for implementing the cyclic POVM for $m=4$ (right side).}
  \label{Figur DFT}
\end{figure}

For the construction of a quantum circuit we have to embed the system
of dimension $m$ into a register with $l$ qubits. The register must
have $r := 2^l \geq m$ dimensions. Following Section \ref{Sektion
ortho} we extend the cyclic POVM by an appropriate number of zero
operators $A_{m+1}, \ldots , A_{r} = 0_2 \in {\mathbb C}^{2 \times
2}$. We therefore have $$M = \sqrt{\frac{1}{m}}\left(
\begin{array}{cccc|ccc} 1 & 1 & \ldots & 1 & 0 & \ldots & 0 \\ 1 &
\omega & \ldots & \omega^{m-1} & 0 & \ldots & 0 \end{array} \right)
\in {\mathbb C}^{2 \times r}.$$ A possible unitary extension ${\tilde
M}$ of this matrix is given by ${\tilde M} = F_m \oplus I_{r-m} \in
{\mathbb C}^{r \times r}$ where $I_{r-m}$ denotes the identity matrix
of size $r-m$.  Consequently, on a qubit register the cyclic POVM
corresponding to the $m$-sided regular polygon can be implemented by
performing the operation ${\tilde M}^\dagger = F_m^\dagger \oplus I_{r
- m}$.  The circuit for implementing the cyclic POVM is schematically
shown in Figure \ref{Figur DFT}. Note that the embedding $\rho \mapsto
\rho \oplus 0_{r-m}$ corresponds to the use of initialized ancilla
qubits.

The Fourier transform can be implemented efficiently if $m$ is a power
of two \cite{Nielsen}. Furthermore, the embedding into a qubit
register is straightforward since we do not need zero operators in
this case.  In summary, the cyclic POVM can be implemented efficiently
on a qubit register if $m$ is a power of two. For instance, the
quantum circuit for the implementation of the cyclic POVM is shown in
Figure \ref{Figur DFT} for $m=4$. The circuit of $F_4^\dagger$ is the
standard circuit for Fourier transforms \cite{Nielsen}.  Note that the
first permutation of the qubits can be removed when we change the
order of the input.

\section{Dihedral groups}
\label{Section 2.2}
The cyclic symmetry group of an $m$-sided regular polygon which we
considered in the previous section is a subgroup of the dihedral
group. The dihedral group consists of all rotations which map the
$m$-sided regular polygon onto itself. In contrast to the cyclic group
we allow the rotations to have different axes. For a fixed $m\geq 2$,
the dihedral group is isomorphic to $D_m=\langle r,s\rangle$ with
$r^m=1$, $s^2=1$, and $srs=r^{-1}$. In order to use the results for
the cyclic groups, we consider the same orientation of the regular
polygon as in the previous section, i.e., the face of the polygon is
orthogonal to the $z$-axis. Furthermore, we assume that at least one
vertex is an element of the $x$-axis. Due to this orientation, the
element $r$ corresponds to the $2 \pi/m$ rotation about the $z$-axis
and the element $s$ corresponds to the $\pi$ rotation about the
$x$-axis.  In the Hilbert space these rotations correspond to the
matrices
$$\left(\begin{array}{cc}1 &0 \\ 0& \omega \end{array} \right) \; {\rm
and} \; \; \left( \begin{array}{cc}0&1\\1&0\end{array}\right)$$ where
$\omega:={\rm exp}(-2\pi i/m)$ is a $m$th complex root  of unity. We can
define a projective representation of the group $D_m$ by mapping the
element $r \in D_m$ to the first matrix and the element $s \in D_m$ to
the second matrix.

\begin{figure}
  \centerline{\epsffile{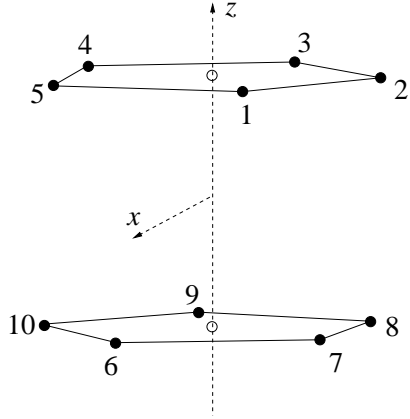}}
  \caption{Points of the dihedral POVM with $m=5$.}
  \label{Figur diederpunkte}
\end{figure}

We consider the orbit of a vector under the action of the dihedral
group $D_m$. Let $(\alpha, \beta)^T \in {\mathbb C}^2$ with
$|\alpha|^2 + |\beta|^2 =1$. Since a global phase factor of a vector
is physically irrelevant we assume $\alpha \in {\mathbb R}$ without
loss of generality. Under the action of the dihedral group the orbit
contains the vectors $(\alpha,\beta \omega^j)^T$ and $(\beta, \alpha
\omega^{j})^T$ with $j \in \{0, \ldots , m-1\}$. An example of the
orbit is shown in Figure \ref{Figur diederpunkte}.
We have at most $n=2m$ vectors. In the following, we assume that the
orbit contains $2m$ elements. If the orbit of $D_m$ contains less than
$2m$ points we have either the case that all points are on the
$xy$-plane 
(and the POVM consists of a single orbit under the group $C_m$)
or we have only the two points $(1,0)^T$ and $(0,1)^T$ defining an
orthogonal measurement. Since
$$\sum_j \left(
\begin{array}{c} \alpha \\ \beta \omega^j \end{array} \right) (\alpha,
{\overline \beta} \omega ^{-j}) + \sum_j \left(
\begin{array}{c} \beta \\ \alpha \omega^j \end{array} \right)
({\overline \beta}, \alpha \omega^{-j}) = m I_2,$$ we rescale $\alpha$
and $\beta$ with the factor $\sqrt{1/m}$ to obtain a POVM.

We now consider the implementation of the dihedral POVM. In order to
analyze the structure, we do not consider the embedding of the
constructed system into a qubit register in the first place. The orbit
under the action of $D_m$ breaks into two orbits under the action of
the subgroup $C_m$.  The two orbits can be obtained by the action of
$C_m$ on the vectors $(\alpha, \beta)^T$ and $(\beta,
\alpha)^T$. Therefore, we expect to obtain implementations of the
dihedral POVMs which are similar to the implementations in the
previous section. With an appropriate order of the vectors we have the
matrix $$M= \left(
\begin{array}{cccc|cccc} \alpha & \alpha & \ldots & \alpha
& \beta & \beta & \ldots & \beta \\ \beta & \beta \omega & \ldots &
\beta \omega^{m-1} & \alpha & \alpha \omega & \ldots & \alpha
\omega^{m-1} \end{array} \right) \in {\mathbb C}^{2 \times 2m}.$$ For
even $m$, this matrix can be extended to the unitary matrix
$${\tilde M} = Q  \left( \begin{array}{cccc|cccc} \alpha &
\alpha & \ldots & \alpha & \beta & \beta & \ldots & \beta \\ \alpha &
\alpha \omega & \ldots & \alpha \omega^{m-1} & -{\overline \beta} &
-{\overline \beta} \omega & \ldots & -{\overline \beta} \omega^{m-1} \\
\alpha & \alpha \omega^2 & \ldots & \alpha \omega^{2(m-1)} & \beta &
\beta \omega^2 & \ldots & \beta \omega^{2(m-1)} \\ \vdots &&& \vdots &
\vdots &&& \vdots \\ \alpha & \alpha \omega^{m-1} & \ldots & \alpha
\omega^{(m-1)(m-1)} & -{\overline \beta} & -{\overline \beta}
\omega^{m-1} & \ldots & -{\overline \beta} 
\omega^{(m-1)(m-1)}  \\ \hline &&&&&&\\ [-0.35cm]
 {\overline \beta} & {\overline \beta} & \ldots & 
{\overline \beta} & -\alpha & -\alpha & \ldots & -\alpha \\ \beta &
\beta \omega & \ldots & \beta \omega^{m-1} & \alpha & \alpha \omega
& \ldots & \alpha \omega^{m-1} \\ {\overline \beta} & {\overline
\beta} \omega^2 & \ldots & {\overline \beta} \omega^{2(m-1)} & -\alpha
& -\alpha \omega^2 & \ldots & -\alpha \omega^{2(m-1)} \\ \vdots &&&
\vdots & \vdots &&& \vdots \\ \beta & \beta
\omega^{m-1} & \ldots & \beta \omega^{(m-1)(m-1)} & 
\alpha & \alpha \omega^{m-1} & \ldots & \alpha \omega^{(m-1)(m-1)} 
\end{array} \right)$$ with a permutation matrix $Q \in {\mathbb C}^{n \times
n}$ fixing the first row and mapping the $(m+2)$nd row to the second
row. For odd $m$, the extended matrix is similar. We only have to
write $\beta \omega^{(m-1)j}$ instead of $- {\overline \beta}
\omega^{(m-1)j}$ in the $m$th row. In the last row we write
${\overline \beta} \omega^{(m-1)j}$ and $-\alpha \omega^{(m-1)j}$
instead of $\beta \omega^{(m-1)j}$ and $\alpha \omega^{(m-1)j}$,
respectively. In order to simplify notation, we mainly consider the
case of even $m$ in the following. The constructions for odd $m$ are
similar.

\begin{figure}
  \centerline{\epsffile{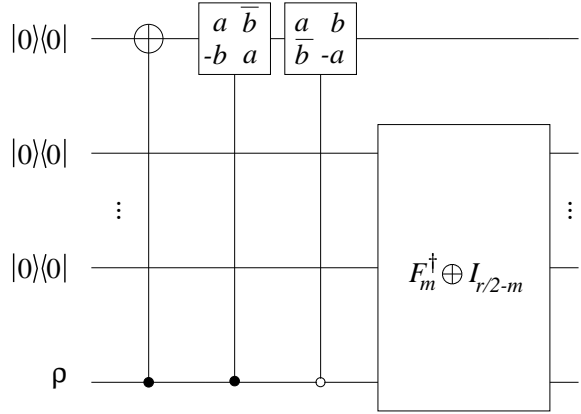}} \caption{A quantum
  circuit for implementing the dihedral POVM. We set
  $a:=\sqrt{m}\alpha$ and $b:=\sqrt{m}\beta$ to simplify notation.}
  \label{Figur schaltkreis dieder-2pt}
\end{figure}

We consider a decomposition of the matrix $Q^\dagger {\tilde M}$
to obtain a decomposition of ${\tilde M}$. The matrix $Q^\dagger {\tilde M}$ 
can be multiplied with
$I_2 \otimes F_m^\dagger$ from the right
leading to $$T= \sqrt{m}\left( \begin{array}{ll} {\rm diag}(\alpha, \alpha,
\alpha, \ldots, \alpha, \alpha ) & {\rm diag}(+\beta, -{\overline
\beta}, +\beta, \ldots, +\beta, -{\overline \beta})
\\ {\rm diag}({\overline \beta}, \beta, {\overline \beta}, \ldots,
{\overline \beta} , \beta) &
{\rm diag}(-\alpha,+ \alpha, -\alpha, \ldots,  -\alpha, + \alpha ) 
\end{array} \right).$$ 
We now embed the system with $n=2m$ dimensions into a qubit register.
We consider a register with $l$ qubits where $r := 2^l \geq n$.
We replace the
matrix $T$ with the matrix $T_r$ of the same structure but of size
$r$. This is done by extending each of the four diagonal components to
a diagonal matrix in ${\mathbb C}^{(r/2) \times (r/2)}$ while conserving
the structure. For instance, the matrix $$T= \sqrt{3}\left(
\begin{array}{ccc|ccc} \alpha &&&\beta\\ &\alpha &&&-{\overline \beta}
\\&&\alpha &&& \beta \\ \hline &&&&\\ [-0.35cm] {\overline \beta} &&&
-\alpha \\ & \beta &&& \alpha \\ && {\overline \beta} &&& -\alpha
\end{array}\right)$$ is extended to the matrix 
$$T_8=\sqrt{3}\left( \begin{array}{cccc|cccc}
\alpha &&&&\beta\\ &\alpha &&&&-{\overline \beta} \\&&\alpha &&&&
\beta \\ &&& \alpha &&&& -{\overline \beta} 
\\ \hline &&&&\\ [-0.35cm] {\overline \beta} &&&& -\alpha \\ & \beta
&&&& \alpha \\ && {\overline \beta} &&&& -\alpha\\ &&& \beta &&&& \alpha
\end{array}\right).$$
Furthermore, in the factorization $T=Q^\dagger {\tilde M} (I_2 \otimes
F_m^\dagger)$ the matrix $Q$ is replaced by a permutation matrix $Q_r
\in {\mathbb C}^{r \times r}$ that fixes the first row and maps the
$(r/2+2)$nd row to the second row. In qubit notation, this permutation
matrix can be described by $\ket{0 \ldots 0} \mapsto \ket{0 \ldots 0}$
and $\ket{10 \ldots 01} \mapsto \ket{00 \ldots 01}$. This permutation
can be implemented by an XOR-gate on the first qubit controlled by the
last qubit. Other implementations that satisfy the two constraints are
also possible. The Fourier transform $F_m$ is replaced by $F_m \oplus
I_{r/2-m}$. In summary, we obtain a matrix ${\tilde M}_r$ that is
defined by the equation
\begin{equation}
\label{Gleichung}
{\tilde M}_r = Q_r T_r (I_2 \otimes (F_m \oplus I_{r/2-m})) \in {\mathbb
C}^{r \times r}.
\end{equation}
This matrix is a unitary extension of the matrix $M$
corresponding to the dihedral POVM with some zero operators as
discussed in Section \ref{Sektion ortho}. Our example with $T_8$ leads
to the matrix
$$Q_8^\dagger {\tilde M}_8 =  \left( 
\begin{array}{cccc|cccc} \alpha & \alpha & \alpha & 0 & \beta & \beta
& \beta & 0 \\ \alpha & \alpha \omega & \alpha \omega^2 & 0 &
-{\overline \beta} & -{\overline \beta} \omega & -{\overline \beta}
\omega^2 & 0 \\ \alpha & \alpha \omega^2 & \alpha \omega & 0 & \beta &
\beta \omega^2 & \beta \omega & 0 \\ 0 & 0 & 0 & \sqrt{3} \alpha & 0 &
0& 0& -\sqrt{3} \, {\overline \beta} \\ 
\hline &&&& \\ [-0.35cm] {\overline \beta} &
{\overline \beta} & {\overline \beta} & 0 & - \alpha & -\alpha &
-\alpha & 0 \\ \beta & \beta \omega & \beta \omega^2 & 0 & \alpha &
\alpha \omega & \alpha \omega^2 & 0 \\ {\overline \beta} & {\overline
\beta} \omega^2 & {\overline \beta} \omega & 0 & - \alpha & - \alpha
\omega^2 & -\alpha \omega & 0 \\ 0 & 0 & 0 & \sqrt{3} \beta & 0 & 0&
0& \sqrt{3}\alpha \end{array}\right).$$ The matrix $Q_8$ maps the
sixth row to the second row leading to the first two rows $$
\left( \begin{array}{cccc|cccc} \alpha & \alpha &
\alpha & 0 & \beta & \beta & \beta & 0 \\ \beta & \beta \omega & \beta
\omega^2 & 0 & \alpha & \alpha \omega & \alpha \omega^2 & 0
\end{array} \right)$$ with two zero columns that do not change the
POVM due to zero probability.  
For convenience, we shift the qubits according to the mapping
$\ket{x_1 \ldots x_{l-1} x_l} \mapsto \ket{x_l x_1 \ldots x_{l-1}}$. We denote
this permutation by $R$. After
this reordering of qubits the matrix $T_r$ takes the simple form
$$RT_rR^\dagger = A \otimes I_{r/4} := 
\sqrt{m}\left( \begin{array}{cccc} \alpha & \beta & 0 &
0 \\ {\overline \beta} & -\alpha & 0 & 0 \\ 0 & 0 & 
\alpha & -{\overline \beta} \\ 0 & 0 & \beta & \alpha 
\end{array} \right) \otimes
I_{r/4} .$$ 
By combining this equation with Equation (\ref{Gleichung}) we get the
factorization $${\tilde M}^\dagger_r = (I_2 \otimes (F_m^\dagger
\oplus I_{r/2-m})) R^\dagger (A^\dagger \otimes I_{r/4}) R
Q_r^\dagger.$$ Translating this equation into a quantum circuit,
the decomposition of ${\tilde M}_r$ leads to the circuit scheme shown
in Figure \ref{Figur schaltkreis dieder-2pt}.
The operation $A^\dagger$ is decomposed as 
$$
A^\dagger= \sqrt{m} \left( \begin{array}{cccc}\alpha &
\beta & 0 & 0 \\ {\overline \beta} & - \alpha & 0 & 0 \\ 0 & 0 &
\sqrt{1/m} & 0 \\ 0 & 0 & 0 & \sqrt{1/m}
\end{array} \right) 
\sqrt{m} \left( \begin{array}{cccc} \sqrt{1/m} & 0
& 0 &0 \\ 0 & \sqrt{1/m} & 0 & 0 \\ 0 & 0 & \alpha & {\overline \beta}
\\ 0 & 0 & - \beta & \alpha \end{array} \right)
$$
corresponding to the second and third gates from the left in Figure
\ref{Figur schaltkreis dieder-2pt}. We do not have to implement the
permutation $R$ explicitly if the controlled one-qubit operations are
applied to appropriate qubit pairs.
The given circuit can be slightly simplified by merging the first two
gates from the left to a single controlled gate with the operation
$$\sqrt{m}\left( \begin{array}{cc} \alpha & {\overline \beta} \\ -
\beta & \alpha\end{array} \right) \left( \begin{array}{cc}0&1\\1&0
\end{array} \right) = \sqrt{m} \left(
\begin{array}{cc}{\overline \beta} & \alpha \\ \alpha & -\beta
\end{array} \right).$$
As discussed in the previous section the Fourier transform $F_m$ can be
implemented with a polylogarithmical number of  gates if $m$ is a power of
two. Consequently, the dihedral POVM can be implemented efficiently in
these cases.

\section{Tetrahedron}
\label{S 4}
\begin{figure}
  \centerline{\epsffile{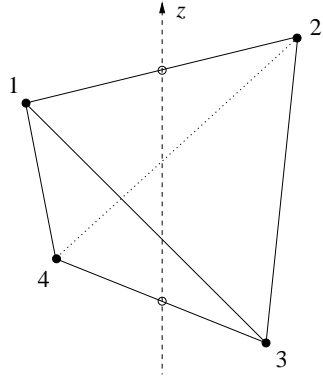}}
  \caption{The tetrahedron with two edges perpendicular to the
  $z$-axis.}
  \label{Figur tetra-schraube}
\end{figure}

The tetrahedron is the platonic solid with four faces. The symmetry
group of the tetrahedron is isomorphic to the alternating group
$A_4$. This group consists of the twelve permutations of four elements
with positive signum. We consider the POVM corresponding to the
vertices of the tetrahedron in the Bloch sphere. The tetrahedron is
shown Figure \ref{Figur tetra-schraube}.
For instance, the vertex $1$ is given by the vector $(\sqrt{2/3},0,
\sqrt{1/3})^T \in {\mathbb R}^3$.
The vertices of the tetrahedron correspond to the vectors
\begin{equation}
\label{Gleichung 2}
\left( \begin{array}{c} \alpha \\ \beta \end{array} \right), \left(
\begin{array}{c} \alpha \\ - \beta \end{array} \right), \left(
\begin{array}{c} \beta \\ \alpha i\end{array} \right), \left(
\begin{array}{c} \beta \\ - \alpha i \end{array} \right) \in {\mathbb C}^2
\end{equation} with
$\alpha = \sqrt{(3+\sqrt{3})/6}$ and $\beta = \sqrt{
(3-\sqrt{3})/6}$. The first pair of vectors corresponds to the
vertices $1$ and $2$, the second pair corresponds to the vertices $3$
and $4$. Note the similarity of these vectors to the vectors
$$\left( \begin{array}{c} \alpha \\ \beta \end{array} \right), \left(
\begin{array}{c} \alpha \\ - \beta \end{array} \right), \left(
\begin{array}{c} \beta \\ \alpha \end{array} \right), \left(
\begin{array}{c} \beta \\ - \alpha  \end{array} \right).$$ These
vectors result from the action of the dihedral group with $m=2$ as
considered in previous section with the vector $(\alpha,
\beta)^T$. The factor $i$ in the second component of the last two
vectors of Line (\ref{Gleichung 2}) results from the $\pi/2$ rotation
about the $z$-axis of the lower edge with vertices 3 and 4 relative to
the upper edge with vertices 1 and 2. This rotation corresponds to the
matrix ${\rm diag}(1,i) \in {\mathbb C}^{2 \times 2}$. Due to the
equation
$$\left(\begin{array}{c} \alpha \\ \beta \end{array} \right) (\alpha,
\beta) + \left(\begin{array}{c} \alpha \\ - \beta \end{array} \right)
(\alpha, -\beta) + \left(\begin{array}{c} \beta \\ \alpha i
\end{array} \right) (\beta, - \alpha i ) + \left(\begin{array}{c}
\beta \\ - \alpha i
\end{array} \right) (\beta, \alpha i) = 2  I_2$$ we have the matrix
$$M=\left(
\begin{array}{cc|cc} \alpha & \alpha & \beta & \beta \\ \beta & -\beta
& \alpha i & - \alpha i \end{array} \right) \in {\mathbb C}^{2 \times
4}$$ with the rescaled elements $\alpha = \sqrt{
(3 + \sqrt{3})/12}$ and $\beta = \sqrt{ (3 -\sqrt{3})/12}$. This
matrix can be extended to the unitary matrix $${\tilde M} = Q  \left(
\begin{array}{cc|cc} \alpha & \alpha & \beta & \beta \\ \alpha & -
\alpha & - \beta \, i & \beta \, i \\ \hline \beta & \beta & - \alpha &
-\alpha \\ \beta & - \beta &  \alpha \,  i & -\alpha \, i \end{array}
\right) \in {\mathbb C}^{4 \times 4}$$ 
acting on a register of two qubits with the permutation matrix $$Q=\left(
\begin{array}{cccc} 1 & 0 & 0 & 0 \\ 0 & 0 & 0 & 1 \\ 0 & 0 & 1 & 0 \\
0 & 1 & 0 & 0 \end{array} \right) \in {\mathbb C}^{4 \times 4}.
$$ The matrix $Q$ can be implemented by an XOR-gate on the first qubit
controlled by the second qubit. We consider the decomposition of
$Q^\dagger {\tilde M}$ to obtain a decomposition of ${\tilde
M}$. After multiplying $Q^\dagger {\tilde M}$ with $(I_2 \otimes F_2)
\in {\mathbb C}^{4 \times 4}$ we have $$Q^\dagger {\tilde M}(I_2
\otimes F_2 ) = \sqrt{2}\left( \begin{array}{cc|cc} \alpha & 0 & \beta
& 0 \\ 0 & \alpha & 0 & - \beta \, i \\ \hline \beta & 0 & -\alpha & 0
\\ 0 & \beta & 0 & \alpha \, i \end{array} \right) \in {\mathbb C}^{4
\times 4}$$ and after multiplying this matrix with ${\rm
diag}(1,1,1,i) \in {\mathbb C}^{4 \times 4}$
\begin{figure}
  \centerline{\epsffile{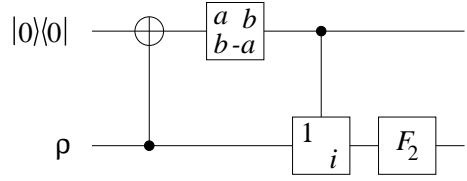}} \caption{The
  circuit for the tetrahedral POVM. We set $a:=\sqrt{2} \alpha$ and
  $b:=\sqrt{2} \beta$ to simplify notation.}
\label{Figur tetra-schraube-kreis}
\end{figure} 
from the right we have the equation
\begin{equation}
\label{Gleichung 3}
Q^\dagger {\tilde M} (I_2 \otimes F_2) \left(
\begin{array}{cccc} 1&0&0&0 \\ 0&1&0&0\\ 0&0& 1&0\\ 0&0&0&i 
\end{array} \right) =
\left( \sqrt{2} \left( \begin{array}{cc} \alpha & \beta \\ \beta & - \alpha
\end{array} \right) \otimes I_2 \right) \in {\mathbb C}^{4 \times
4}. \end{equation} The matrix ${\rm diag}(1,1,1,i)$ corresponds to a
controlled phase gate ${\rm diag}(1,i)$ on the second qubit. 
Using Equation (\ref{Gleichung 3}) we get the equation $${\tilde M}^\dagger =
(I_2 \otimes F_2) \left(
\begin{array}{cccc}1&0&0&0\\0&1&0&0\\0&0&1&0\\0&0&0&i\end{array}\right) 
\left( \sqrt{2} \left(
\begin{array}{cc} \alpha & \beta \\ \beta & - \alpha \end{array}
\right) \otimes I_2 \right) Q^\dagger.$$
Consequently, the circuit in Figure \ref{Figur tetra-schraube-kreis}
implements the transformation ${\tilde M}^\dagger$ for the POVM
corresponding to the tetrahedron.

\section{Cube}
\label{S 6}
\begin{figure}
  \centerline{\epsffile{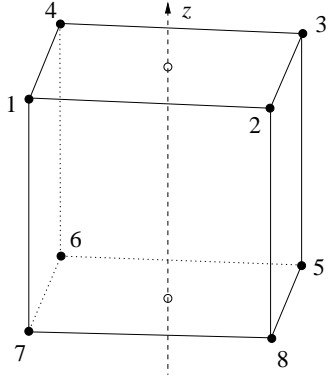}}
  \caption{The cube with two faces perpendicular to the $z$-axis.}
  \label{Figur cube}
\end{figure}

\begin{figure}
  \centerline{\epsffile{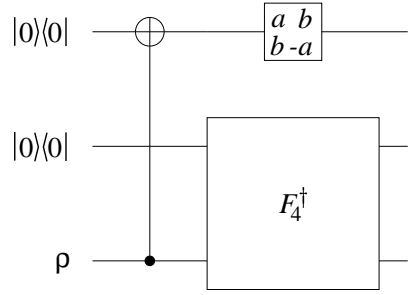}} \caption{The circuit
  implementing the cubic POVM. We set $a:=2\alpha$ and $b:=2\beta$ to
  simplify notation.}
\label{Figur Cube-speciale-kreis}
\end{figure}

The POVM associated with a cube in the Bloch sphere is a special case
of the dihedral POVMs considered in Section \ref{Section 2.2} with
$m=4$. Nevertheless, we consider the implementation of the cubic POVM
in this section since we can obtain a smaller circuit by using the
special values of $\alpha$ and $\beta$.
As in Section~\ref{Section 2.2} we rotate the cube in the Bloch sphere
to obtain a face perpendicular to the $z$-axis. Furthermore, we can
rotate the cube about this axis to get points corresponding to the
vectors
$$\left( \begin{array}{c} \alpha \\ \beta \end{array} \right), \left(
\begin{array}{c} \alpha \\ \beta i \end{array} \right), \left(
\begin{array}{c} \alpha \\ - \beta \end{array} \right), \left(
\begin{array}{c} \alpha \\ - \beta i \end{array} \right), \left(
\begin{array}{c} \beta \\ - \alpha \end{array} \right), \left(
\begin{array}{c} \beta \\ - \alpha i \end{array} \right), \left(
\begin{array}{c} \beta \\ \alpha \end{array} \right), \left(
\begin{array}{c} \beta \\  \alpha i\end{array} \right) \in {\mathbb
C}^2$$ with 
$\alpha=\sqrt{(3+\sqrt{3})/6}$ and $\beta = \sqrt{(3-\sqrt{3})/6}$.
The first
four vectors correspond to vertices 1--4 in Figure \ref{Figur cube}
and
the last four vectors correspond to vertices 5--8. 
For instance, the vertex $1$ corresponds to the Bloch point $ 
(\sqrt{2/3},0,\sqrt{1/3})^T \in {\mathbb R}^3$.
Note that
$\alpha$ and $\beta$ are real numbers. This allows us to use a more
efficient construction than in Section \ref{Section 2.2}. Since we
have the equation 
$$ \left(\begin{array}{c}\alpha\\ \beta \end{array} \right) ( \alpha,
\beta) + \ldots + \left(\begin{array}{c}\alpha\\ - \beta i\end{array}
\right) ( \alpha,  \beta i) + \left(\begin{array}{c}\beta\\ -\alpha
\end{array} \right) ( \beta, -\alpha) + \ldots +
\left(\begin{array}{c}\beta\\ \alpha i \end{array} \right) ( \beta, -
\alpha i) = 4  I_2$$ the given vectors define a POVM when we
rescale $\alpha$ and $\beta$ with $1/2$. The matrix $M$ corresponding
to the POVM is given by $$M = \left(
\begin{array}{cccc|cccc}\alpha & \alpha & \alpha & \alpha & \beta &
\beta & \beta & \beta \\ \beta & \beta i & - \beta & - \beta i & -
\alpha & - \alpha i & \alpha & \alpha i \end{array} \right) \in
{\mathbb C}^{2 \times 8}.$$ In contrast to the construction of the
dihedral POVM we use the fact that for even $m$ the element $-1$ is in
the set $\{1, \omega,\ldots, \omega^{m-1} \}$ where $\omega:={\rm
exp}(-2\pi i/m)$ is an $m$-th complex root of unity. Therefore, we can
reorder the vectors $({\overline \beta}, \alpha \omega^j)^T \in
{\mathbb C}^2$ considered in Section \ref{Section 2.2} to obtain a
matrix $M$ with the partial row $(-\alpha, -\alpha \omega, \ldots,
-\alpha \omega^{m-1})$ instead of $(\alpha, \alpha \omega, \ldots,
\alpha \omega^{m-1})$. This is besides $\beta \in {\mathbb R}$ the
second reason that allows a more efficient construction compared to
the construction for the dihedral POVM.  Using the equation $\beta =
{\overline \beta}$ we can extend $M$ to the unitary matrix
$${\tilde M} = Q \left( \begin{array}{cccc|cccc} 
\alpha & \alpha & \alpha & \alpha & \beta & \beta & \beta & \beta \\ 
\alpha & \alpha i& -\alpha & -\alpha i& \beta & \beta i& -\beta & -\beta i\\ 
\alpha & -\alpha & \alpha &-\alpha & \beta & -\beta & \beta & -\beta \\ 
\alpha & -\alpha i& -\alpha & \alpha i& \beta & -\beta i& -\beta & \beta i\\ 
\hline \beta&\beta&\beta&\beta&-\alpha&-\alpha&-\alpha&-\alpha\\
\beta&\beta i&-\beta&-\beta i&-\alpha&-\alpha i&\alpha&\alpha i\\
\beta&-\beta&\beta&-\beta&-\alpha&\alpha&-\alpha&\alpha\\
\beta&-\beta i& -\beta&\beta i&-\alpha&\alpha i&\alpha&-\alpha i
\end{array}\right)$$ acting on a register of three qubits
with a permutation $Q$ satisfying $\ket{000} \mapsto \ket{000}$ and
$\ket{101} \mapsto \ket{001}$ in qubit notation. For instance, the
permutation can be implemented by a single XOR-gate on the first qubit
controlled by the third.  We now consider the special structure of
$Q^\dagger {\tilde M}$ to obtain a decomposition of ${\tilde
M}^\dagger$. More precisely, the matrix $Q^\dagger {\tilde M}$ can be
written as the following tensor product $$ Q^\dagger {\tilde M} =
\left( 2\left(
\begin{array}{cc}\alpha & \beta \\ \beta & - \alpha \end{array}
\right) \otimes F_4\right) \in {\mathbb C}^{8 \times 8}.$$ This
leads to the identity
$${\tilde M}^\dagger = 
\left( 2 \left( \begin{array}{cc}\alpha & \beta \\ \beta & - \alpha
\end{array} \right) \otimes F_4^\dagger \right) Q^\dagger
\in {\mathbb C}^{8 \times 8}$$ defining the quantum circuit 
given in Figure \ref{Figur Cube-speciale-kreis}. 
Compared to the general circuit in Section \ref{Section 2.2} we are
able to replace two controlled gates by a single uncontrolled gate.

\section{Octahedron}
\label{S Okta}
\begin{figure}
  \centerline{\epsffile{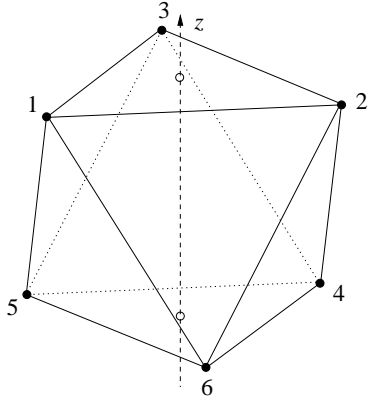}}
  \caption{The octahedron with two faces perpendicular to the $z$-axis.}
  \label{Figur Oktaeder}
\end{figure}

The symmetry group of the octahedron is identical to the symmetry
group of the cube since the octahedron is the dual polyhedron of the
cube. The group is isomorphic to the symmetric group $S_4$. This group
consists of all $24$ permutations of four elements.  A simple
implementation of the octahedral POVM can be obtained by the
orientation of the octahedron as shown in Figure \ref{Figur Oktaeder}
where the upper face with vertices 1--3 and the lower face with
vertices 4--6 are perpendicular to the $z$-axis. Vertex $1$
corresponds to the real vector $(\sqrt{2/3},0,\sqrt{1/3})^T \in
{\mathbb R}^3$.  The complex vectors corresponding to the points 1--6
are given by
$$
\left(\begin{array}{c} \alpha \\ \beta \end{array} \right),
\left(\begin{array}{c} \alpha \\ \beta \omega \end{array} \right),
\left(\begin{array}{c} \alpha \\ \beta \omega^2 \end{array} \right),
\left( \begin{array}{c} \beta \\ - \alpha \end{array}\right), \left(
\begin{array}{c} \beta \\ - \alpha \omega \end{array}\right), \left(
\begin{array}{c} \beta \\ - \alpha \omega^2 \end{array} \right) \in
{\mathbb C}^2$$
where $\omega:= {\rm exp}(-2\pi i /3)$ is 
a root  of unity, 
$\alpha = \sqrt{(3+\sqrt{3})/6}$ and $\beta =
\sqrt{(3-\sqrt{3})/6}$.
The first three vectors
correspond to the upper three vertices 1--3 of the octahedron, the
last three vectors to the lower three vertices 4--6. Despite the
negative sign of the second component of the last three elements,
these vectors are identical with the vectors
$$\left(\begin{array}{c} \alpha \\ \beta \end{array} \right),
\left(\begin{array}{c} \alpha \\ \beta \omega \end{array} \right),
\left(\begin{array}{c} \alpha \\ \beta \omega^2 \end{array} \right),
\left( \begin{array}{c} \beta \\ \alpha \end{array}\right), \left(
\begin{array}{c} \beta \\ \alpha \omega \end{array}\right), \left(
\begin{array}{c} \beta \\ \alpha \omega^2 \end{array} \right)
\in {\mathbb C}^2.$$ The latter vectors are obtained by the vector
$(\alpha, \beta)^T$ under the action of the dihedral group $D_3$ as
discussed in Section \ref{Section 2.2}. Similar to the factor $i$ in
two vectors of the tetrahedral POVM in Section \ref{S 4}, the negative
sign results from the $\pi$ rotation about the $z$-axis of the lower
three vertices 4--6 relative to the upper three vertices 1--3.  Since
$$\left( \begin{array}{c} \alpha \\ \beta
\end{array} \right) (\alpha, \beta) + \ldots + \left(\begin{array}{c}
\beta \\ -\alpha \omega^2 \end{array} \right) (\beta, -\alpha \omega
)= 3 I_2$$ we rescale $\alpha$ and $\beta$ with the factor
$\sqrt{1/3}$ to obtain a POVM.  Therefore, we have
\begin{equation}
\label{E 3}
M = \left(
\begin{array}{ccc|ccc}\alpha & \alpha & \alpha & \beta & \beta & \beta
\\ \beta & \beta \omega & \beta \omega^2& - \alpha & - \alpha \omega &
- \alpha \omega^2 \end{array}\right) \in {\mathbb C}^{2 \times 6}.
\end{equation}
As already discussed, we have the negative signs in Equation (\ref{E 3})
because the lower face is rotated relatively to the upper face.  This
is different from the cubic POVM where we have to reorder the
operators (compared to the dihedral case) in order to get the negative
signs in the second component of the last four vectors. To see this,
we write these components as $-\alpha$, $-\alpha \omega$, $- \alpha
\omega^2$ and $-\alpha \omega^3$ with $\omega =i$ as in the dihedral
case.


We now consider the extension of $M$ to a unitary matrix
${\tilde M}$. As in Section \ref{Section 2.2} we do not embed
the system with six dimensions into a qubit register in the first place. 
The matrix $M$
corresponds to the first two rows of the matrix $$ {\tilde M} = Q
\left( \begin{array}{ccc|ccc} \alpha & \alpha & \alpha & \beta &
\beta & \beta \\ \alpha & \alpha \omega & \alpha \omega^2 & \beta &
\beta \omega & \beta \omega^2 \\ \alpha & \alpha \omega^2 & \alpha
\omega & \beta & \beta \omega^2 & \beta \omega \\ \hline \beta & \beta
& \beta & -\alpha & - \alpha & - \alpha \\ \beta & \beta \omega &
\beta \omega^2 & -\alpha & - \alpha \omega & - \alpha \omega^2 \\
\beta & \beta \omega^2 & \beta \omega & -\alpha & - \alpha \omega^2 &
- \alpha \omega \end{array} \right)$$ where $Q$ is a permutation
matrix that fixes the first row and maps the fifth row to the
second. Similar to the previous section, this matrix can be written as 
\begin{equation}
\label{E 10}
{\tilde M} = Q \left( \sqrt{3}
\left( \begin{array}{cc}
\alpha & \beta \\ \beta & - \alpha \end{array} \right) \otimes
F_3\right) \in {\mathbb C}^{6 \times 6}.
\end{equation}

We now translate the decomposition of ${\tilde M}$ into a circuit. We
have to embed the system with six dimensions into a qubit register
with at least three qubits. This can be done by replacing the
Fourier matrix $F_3$ in Equation (\ref{E 10}) with $F_3 \oplus I_1$
where $I_1 \in {\mathbb C}^{1\times 1}$ denotes the identity
matrix of size one. 
This replacement leads to the matrix
$${\tilde M}_8 = Q_8 
\left( \begin{array}{cccc|cccc} \alpha & \alpha & \alpha
& 0 & \beta & \beta & \beta & 0 \\ \alpha & \alpha \omega & \alpha
\omega^2 & 0 & \beta & \beta \omega & \beta \omega^2 & 0 \\ \alpha &
\alpha \omega^2 & \alpha \omega & 0 & \beta & \beta \omega^2 & \beta
\omega & 0 \\ 0 & 0 & 0 & \sqrt{3} \alpha & 0 & 0 & 0 & \sqrt{3} \beta 
\\ \hline \beta & \beta & \beta
& 0 & -\alpha & -\alpha & -\alpha & 0 \\ \beta & \beta \omega & \beta
\omega^2 & 0 & -\alpha & -\alpha \omega & -\alpha \omega^2 & 0 \\ \beta &
\beta \omega^2 & \beta \omega & 0 & -\alpha & -\alpha \omega^2 & -\alpha
\omega & 0 \\ 0 & 0 & 0 & \sqrt{3} \beta & 0 & 0 & 0 & - \sqrt{3} \alpha
\end{array} \right)$$ where $Q_8 \in {\mathbb C}^{8 \times 8}$ 
is a permutation matrix that fixes the first row and maps the sixth
\begin{figure}
  \centerline{\epsffile{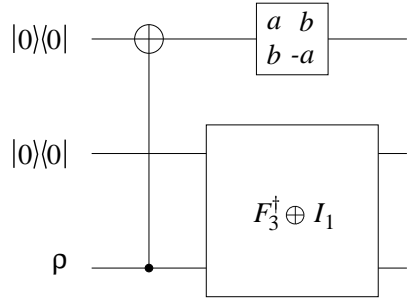}} \caption{A circuit for
  implementing the octahedral POVM. We set $a:=\sqrt{3}\alpha$ and
  $b:=\sqrt{3}\beta$ to simplify notation.}
\label{Figur Kreis Okta}
\end{figure}
row to the second row. In qubit notation, these constraints are given
by $\ket{000} \mapsto \ket{000}$ and $\ket{101} \mapsto
\ket{001}$. For instance,  this transformation can be implemented by 
an XOR-gate on the first qubit controlled by the last qubit. 
If we restrict ${\tilde M}_8$ to the first two rows we get the matrix
$$\left( \begin{array}{cccc|cccc} \alpha & \alpha & \alpha & 0 &
\beta & \beta & \beta & 0 \\ \beta & \beta \omega & \beta \omega^2 &
0 & -\alpha & -\alpha \omega & - \alpha \omega^2 & 0 \end{array}
\right)$$ corresponding to the desired POVM.  
The POVM operator corresponding to the fourth and eighth
column is $0_2 \in {\mathbb C}^{2 \times 2}$ leading to a zero
probability for all states $\rho \in {\mathbb C}^{2 \times 2}$. 
In summary, we have the equation $$ 
{\tilde M}_8^\dagger = \left( \sqrt{3} \left(
\begin{array}{cc} \alpha & \beta \\ \beta & - \alpha \end{array}
\right) \otimes \left( F_3^\dagger \oplus I_1 \right) \right) 
Q_8^\dagger$$ for the implementation of the octahedral POVM. 
This equation corresponds to the circuit shown in
Figure \ref{Figur Kreis Okta}.

\section{Dodecahedron}
\begin{figure}
  \centerline{\epsffile{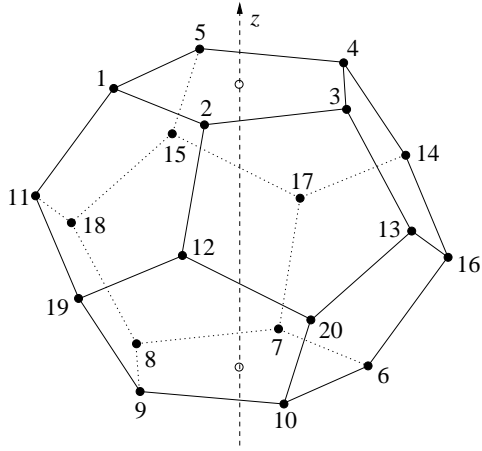}}
  \caption{The dodecahedron with two faces perpendicular to the
  $z$-axis.}
  \label{Figur Dode}
\end{figure}

The dodecahedron is the platonic solid with twelve faces and twenty
vertices.  The symmetry group of the dodecahedron is isomorphic to the
alternating group $A_5$. This group contains the sixty permutations of
five elements with positive signum. The dodecahedron is shown in
Figure \ref{Figur Dode}. The upper face with vertices 1--5 and the
lower face with vertices 6--10 are perpendicular to the $z$-axis.  The
point $$\left( \sqrt{\frac{10-2\sqrt{5}}{15}}\,, \;0 \,, \;
\sqrt{\frac{5+2\sqrt{5}}{15}}\right)^T \in {\mathbb R}^3$$ corresponds
to vertex $1$.  This orientation of the dodecahedron in the Bloch
sphere leads to a simple construction of the dodecahedral POVM.  With
$\omega:={\rm exp}(-2\pi i /5)$, the points on the
Bloch sphere correspond to the complex vectors
\begin{equation}
\label{E 11}
\left(\begin{array}{c}\alpha \\ \beta \omega^j \end{array} \right),
\left(\begin{array}{c}\beta \\ -\alpha \omega^j \end{array} \right),
\left(\begin{array}{c}\gamma \\ \delta \omega^j \end{array} \right),
\left(\begin{array}{c}\delta \\ -\gamma \omega^j \end{array} \right)
\in {\mathbb C}^2\end{equation} where $j \in \{0,\ldots,4\}$.  The
vectors $(\alpha, \beta \omega^j)^T$ correspond to the points 1--5,
the vectors $(\beta, -\alpha \omega^j)^T$ to 6--10, the vectors
$(\gamma, \delta \omega^j)^T$ to 11--15, and the vectors $(\delta,
\gamma \omega^j)^T$ to the points 16--20.  The parameters $\alpha,
\beta, \gamma,$ and $\delta$ are defined as follows:
$$\alpha = \sqrt{ \frac{1}{2} +\frac{1}{30} \sqrt{75+30\sqrt{5}}}
\, , \; \beta = \sqrt{\frac{1}{2}-\frac{1}{30}\sqrt{75+30\sqrt{5}}}
$$ and $$\gamma = \sqrt{\frac{1}{2} + \frac{1}{30}
\sqrt{75-30\sqrt{5}}} \, , \; \delta =
\sqrt{\frac{1}{2}-\frac{1}{30}\sqrt{75-30\sqrt{5}}} .$$ Due to the
equation $$\left(
\begin{array}{c} \alpha \\ \beta \end{array} \right)(\alpha, \beta) +
\ldots + \left(\begin{array}{c}\delta \\ \gamma \omega^4 \end{array}
\right) (\delta, \gamma \omega^{-4} )= 10 I_2$$ we rescale the
elements $\alpha, \beta, \gamma, \delta$ with the factor $\sqrt{1/10}$
to obtain a POVM. In contrast to the constructions of Sections
\ref{Section 2.2}--\ref{S Okta} the points on the Bloch sphere
decompose into four different orbits under the rotation about the
$z$-axis. Note that there are two pairs of orbits. In Line (\ref{E
11}) the vectors $(\alpha, \beta \omega^j)^T$ and $(\beta, -\alpha
\omega^j)^T$ are similar to the vectors $(\alpha, \beta \omega^j)^T$
and $(\beta, \alpha \omega^j)^T$. The latter vectors are the orbit of
$(\alpha, \beta)^T$ under the dihedral group with $m=5$ as considered
in Section \ref{Section 2.2}. As in the previous section a $\pi$
rotation of one orbit relative to the other orbit causes the negative
sign of the elements $(\beta, -\alpha \omega^j)^T$. Analogously, the
second orbit under $D_5$ is defined by the third and fourth type of
vectors in Line (\ref{E 11}). In summary, the vertices of the
dodecahedron correspond to two orbits under the dihedral group $D_5$
with a $\pi$ rotation about the $z$-axis of some points on each
orbit. Consequently, we can expect to use a similar construction as in
the previous sections.

\begin{figure}
  \centerline{\epsffile{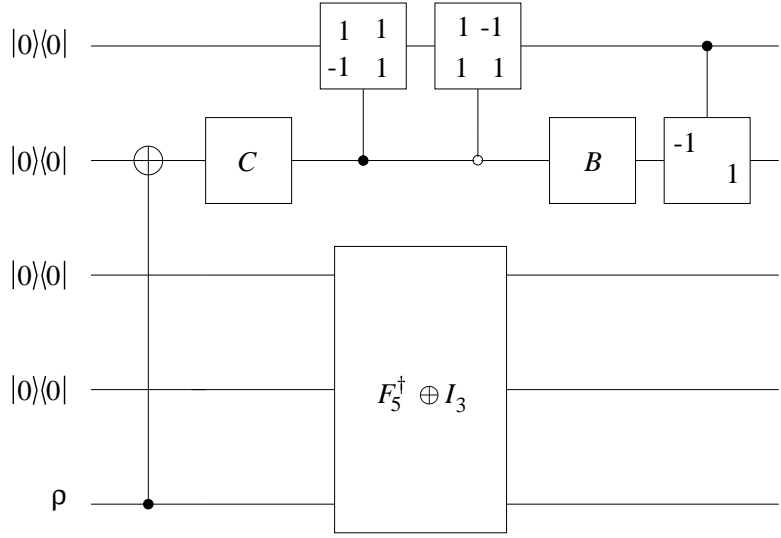}} \caption{A circuit for
  implementing the dodecahedral POVM. In order to simplify notation, the
  elements $\pm 1$ of the gates on the first qubit represent $\pm
  \sqrt{1/2}$.  } \label{Figur Kreis Dodeka}
\end{figure}

We now consider the construction of the circuit for implementing the
dodecahedral POVM. For convenience, we do not embed the system into a
qubit register in the first place.  We have the matrix $$M = \left(
\begin{array}{cccc|ccc|ccc|cccc}\alpha & \alpha & \ldots & \alpha &
\beta & \ldots & \beta & \gamma & \ldots & \gamma & \delta & \delta &
\ldots & \delta \\ \beta & \beta \omega & \ldots & \beta \omega^4 &
-\alpha & \ldots & - \alpha \omega^4 & \delta & \ldots & \delta
\omega^4 & - \gamma & - \gamma \omega & \ldots & - \gamma \omega^4
\end{array} \right).$$ 
This matrix corresponds to the first and second row of the unitary
matrix ${\tilde M}$ defined by the equation
\begin{equation}
\label{E d}
{\tilde M} = Q \left( A 
\otimes F_5 \right) \in {\mathbb C}^{20\times 20},
\end{equation}
where $Q \in {\mathbb C}^{20 \times 20}$ is a permutation matrix that
fixes the first row and maps the seventh row to the second row.  
The matrix $A$ is defined by $$A = \sqrt{5} \left(
\begin{array}{cccc} \alpha & \beta & \gamma & \delta \\ \beta & -
\alpha & \delta & - \gamma \\ \gamma & - \delta & - \alpha & \beta \\
\delta & \gamma & - \beta & - \alpha \end{array} \right).$$ Now, we
want to embed the extended system into a register with five
qubits. Similar to the construction for the octahedron in Section
\ref{S Okta}, we can do this by replacing the matrix $F_5$ in Equation
(\ref{E d}) by the matrix $(F_5 \oplus I_3) \in {\mathbb C}^{8 \times
8}$ where $I_3$ denotes the identity matrix of size three. The matrix
$Q$ is replaced by a permutation matrix $Q_{32} \in {\mathbb C}^{32
\times 32}$ that satisfies $\ket{00000} \mapsto \ket{00000}$ and
$\ket{01001} \mapsto \ket{00001}$ in the qubit notation. This
permutation can be implemented by an XOR-operation on the second qubit
controlled by the last qubit. In summary, the matrix ${\tilde
M}^\dagger_{32}$ is defined by the equation $${\tilde M}^\dagger_{32}
= \left( A^\dagger
\otimes \left( F_5^\dagger \oplus I_3 \right) \right) Q_{32}^\dagger.$$
The corresponding circuit is shown in Figure \ref{Figur Kreis Dodeka}.
Note that the matrix $A^\dagger$ can be written as product 
$A^\dagger = \left(I_2 \oplus (-\sigma_z) \right) \left( I_2 \otimes
B \right) R \left( I_2 \otimes C \right)$ with the matrices 
$$ B= \left( \begin{array}{cc}u_- & -u_+ \\ u_+ &
u_-\end{array}\right),\;\;  C= \left( \begin{array}{cc}
v_-&v_+\\ v_+&-v_-\end{array}\right)$$ and constants 
$$u_\pm = \sqrt{\frac{1}{2} \pm \sqrt{\frac{3+\sqrt{5}}{24}}}\; , \;\;
v_\pm = \mp \sqrt{\frac{1}{2} \pm \sqrt{\frac{\sqrt{5}-1}{8\sqrt{5}}}}.$$
The matrix $R$ is the product
$$R= \sqrt{\frac{1}{2}} \left( 
\begin{array}{cccc}1&0&-1&0\\ 0&\sqrt{2}&0&0 \\ 1&0&1&0 \\ 0&0&0&\sqrt{2}
\end{array}\right) \sqrt{\frac{1}{2}}
\left( \begin{array}{cccc}\sqrt{2}&0&0&0\\ 0&1&0&1 \\ 0&0&\sqrt{2}&0
\\ 0 & -1 & 0 & 1\end{array} \right).$$ In Figure \ref{Figur Kreis
Dodeka}, the latter two matrices correspond to the two operations on
the first qubit which are controlled by the second qubit.


\section{Icosahedron}
\label{S ico}
\begin{figure}
  \centerline{\epsffile{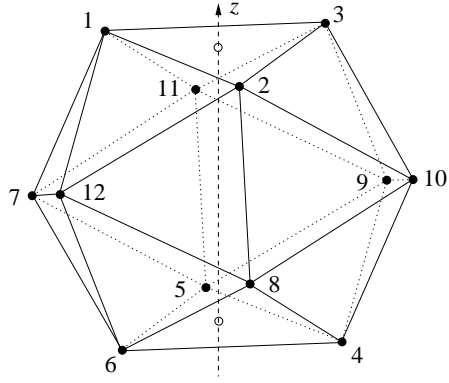}}
  \caption{The icosahedron with two faces perpendicular to the
  $z$-axis.}
 \label{Figur Ikos}
\end{figure}

The icosahedron is the dual polyhedron of the
dodecahedron. Consequently, the symmetry groups of both platonic
solids are identical. We assume the specific orientation of the
icosahedron as shown in Figure \ref{Figur Ikos} to obtain a simple
construction of the icosahedral POVM. The upper face with vertices
1--3 and the lower face with vertices 4--6 are perpendicular to the
$z$-axis. Vertex 1 is given by the vector
$$\left( \sqrt{ \frac{10-2\sqrt{5}}{15}}\,,\; 0 \,,\;
\sqrt{\frac{5+2\sqrt{5}}{15}} \right)^T \in {\mathbb R}^3.$$
The vertices of the icosahedron in the Bloch sphere correspond to the
complex vectors
\begin{equation}
\label{E I}
\left(\begin{array}{c}\alpha \\ \beta \omega^j \end{array}\right), 
\left(\begin{array}{c}\beta \\ - \alpha \omega^j \end{array}\right),
\left(\begin{array}{c}\gamma \\  \delta \omega^j \end{array}\right),
\left(\begin{array}{c}\delta \\ - \gamma \omega^j \end{array}\right)
\in {\mathbb C}^2
\end{equation}
with $j \in \{0,1,2\}$, $\omega:={\rm exp}(- 2 \pi i /3)$
and $$\alpha=\sqrt{\frac{1}{2} + \frac{1}{30} 
\sqrt{ 75 + 30 \sqrt{5}} } \; \; , \; \; \beta = \sqrt{ \frac{1}{2} -
\frac{1}{30} \sqrt{75 + 30 \sqrt{5}} }$$ and
$$\gamma = \sqrt{ \frac{1}{2} - \frac{1}{30}\sqrt{75-30 \sqrt{5}}} \;
\; , \; \; \delta = \sqrt{ \frac{1}{2} + \frac{1}{30}\sqrt{75-30
\sqrt{5}}}.$$ The vectors in Line (\ref{E I}) with $j=0$ correspond to
the vertices $1$, $4$, $7$ and $10$ in the given order. As in the
case of the dodecahedron we have four orbits under the rotations about
the $z$-axis. Therefore, we can expect that a similar construction
as in the previous section is possible. Due to the identity $$ \left(
\begin{array}{c} \alpha \\ \beta \end{array} \right)(\alpha, \beta) +
\ldots + \left(\begin{array}{c}\delta \\ \gamma \omega^4 \end{array}
\right) (\delta, \gamma \omega^{-4} )= 6  I_2$$ we have the matrix
$$M =  \left( \begin{array}{ccc|ccc|ccc|ccc}
\alpha & \alpha & \alpha & \beta & \beta & \beta & \gamma & \gamma &
\gamma & \delta & \delta & \delta \\ \beta & \beta \omega & \beta
\omega^2 & -\alpha & - \alpha \omega & - \alpha \omega^2 & \delta &
\delta \omega & \delta \omega^2 & -\gamma & -\gamma \omega & - \gamma
\omega^2 \end{array} \right),$$ 
where we rescale $\alpha, \beta, \gamma$ and $\delta$ with the factor
$\sqrt{1/6}$. The matrix $M$ consists of the first and second row
of the matrix 
\begin{equation}
\label{EE I}
{\tilde M} = Q  \left( A
\otimes F_3\right) \in {\mathbb C}^{12 \times 12},
\end{equation}
where $Q \in {\mathbb C}^{12 \times 12}$ is a permutation matrix that
fixes the first row and maps the fifth row to the second. Similar to
the previous section, the matrix
$A$ is given by $$A=\sqrt{3}\left( \begin{array}{cccc} \alpha & \beta &
\gamma & \delta \\ \beta & - \alpha & \delta & - \gamma \\ \gamma & -
\delta & - \alpha & \beta \\ \delta & \gamma & - \beta & - \alpha
\end{array} \right).$$

\begin{figure}
  \centerline{\epsffile{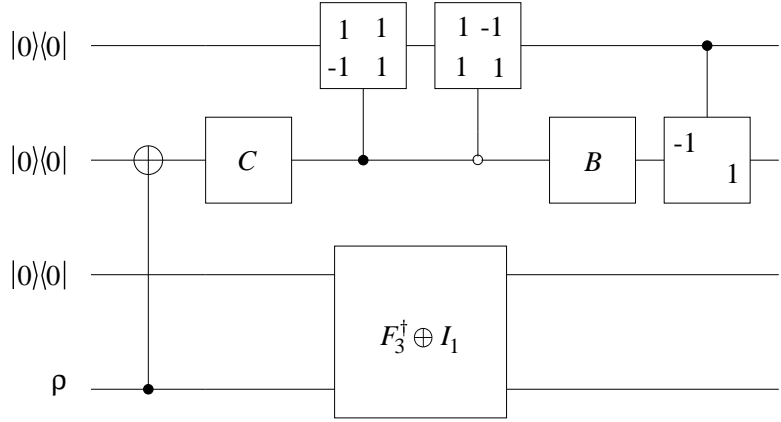}} \caption{A circuit for
  implementing the icosahedral POVM. In order to simplify notation, the
  elements $\pm 1$ of the gates on the first qubit represent $\pm
  \sqrt{1/2}$. } \label{Figur Kreis Ikosa}
\end{figure}

The embedding into a register with four qubits works analogously to
the previous section.  We replace $F_3$ in Equation (\ref{EE I}) by
$(F_3 \oplus I_1) \in {\mathbb C}^{4 \times 4}$ where $I_1$ denotes
the identity matrix of size one. The matrix $Q$ is replaced by the
matrix $Q_{16} \in {\mathbb C}^{16 \times 16}$ that can be described
as $\ket{0000} \mapsto \ket{0000}$ and $\ket{0101} \mapsto \ket{0001}$
in the qubit notation. This permutation can be implemented by an
XOR-operation on the second qubit controlled by the last qubit.
Therefore, Equation (\ref{EE I}) translates into
$${\tilde M}_{16}^\dagger = \left( A^\dagger
\otimes \left( F_3^\dagger \oplus I_1
\right) \right) Q_{16}^\dagger.$$ The circuit corresponding to this
decomposition of ${\tilde M}^\dagger_{16}$ is given in Figure
\ref{Figur Kreis Ikosa}.
The matrix $A^\dagger$ can be translated into single- and two-qubit
gates as shown in the previous section. In this translation we have to
replace the constants $u_\pm$ and $v_\pm$ with 
$$u_\pm = \frac{1}{10} \sqrt{50\pm 5\sqrt{10(5+\sqrt{5})}}
 \quad {\rm and} \quad v_\pm = \mp \frac{1}{2} 
\sqrt{2 \pm \sqrt{5/3} \mp \sqrt{1/3}}.$$

\section{Conclusions}
We have shown that all POVMs given by the vertices of platonic solids
can be implemented using a discrete Fourier transform and a few other
operations. The algorithms use the symmetry of the POVMs.  A common
feature of all constructions is the partition of the POVM operators
into orbits under the action of a cyclic group.  Since the Fourier
transform allows to implement POVMs associated with an orbit under a
cyclic group it is an essential part of all circuits.  For most
POVMs corresponding to a platonic solid, a tensor product of a
Fourier transform and a specific low-dimensional matrix is a central
building block of the circuit. The low-dimensional matrix represents
in some sense the relations between the orbits.

The implementation of non-symmetric POVMs seems to be a non-trivial
task. It would, for instance, be interesting to know which POVMs can
be implemented efficiently, i.e., with a number of elementary gates
which grows only polynomially in the number $n$ of POVM-operators.
For the symmetric POVMs considered in this paper the question of
efficiency makes only sense for the cyclic and dihedral POVMs since
the size of the other POVMs is fixed.  For $n=2^l$ the complexity of
the Fourier transform $F_n$ is only polynomial in $l$. Therefore, the
complexity of the circuits for the cyclic and dihedral POVMs grows
only polylogarithmically in $n$.

The question of the efficiency of read-out mechanisms for a single bit
has no counterpart in classical computer science.  Complexity issues
in quantum information theory deal not necessarily with the complexity
of {\it computational} problems. They are also interesting in the
context of measurements or state preparation procedures. However,
there are some connections between a complexity theory of these
non-computational quantum control problems and computational problems
\cite{FQMA,PSPACE}. Connections between the complexity of POVM
measurements and other complexity issues may be subject of further
research.

The authors acknowledge helpful discussions with M. Grassl and M.
R\"otteler. M. R\"otteler brought the problem of implementing
symmetric POVMs to our attention. 
This work was supported by grants of the BMBF project
01/BB01B.


\begin{thebibliography}{}

\bibitem{Davies2}E.B. Davies: {\it Quantum theory of open systems},
Academic Press, 1976.

\bibitem{Sasaki}M. Sasaki, S.M. Barnett, R. Jozsa, M. Osaki,
O. Hirota: {\it Accessible information and optimal strategies for real
symmetrical quantum sources}, Physical Review A, Vol. 59, No. 5,
pp. 3325--3335, May 1999.

\bibitem{Davies}E.B. Davies: {\it Information and Quantum
Measurement}, IEEE Transactions on Information Theory, Vol. IT-24,
No. 5, September 1978.

\bibitem{Eldar}Y.C. Eldar, G.D. Forney: {\it On Quantum Detection and
the Square-Root Measurement}, IEEE Transactions on Information Theory,
Vol. 47, pp. 858-872, Mar. 2001.

\bibitem{Fuchs} C. Fuchs: {\it Information Gain vs. State Disturbance
in Quantum Theory}, LANL-preprint quant-ph/9611010.

\bibitem{Peres}A. Peres: {\it Quantum Theory: Concepts and Methods},
Kluwer Academic Publishers, 1993.

\bibitem{Nielsen}M.A. Nielsen, I.L. Chuang: {\it Quantum Computation
and Quantum Information}, Cambridge University Press, 2000.

\bibitem{Sternberg}S. Sternberg: {\it Group theory and physics},
Cambridge University Press, 1994.

\bibitem{FQMA}D. Janzing, P. Wocjan, Th. Beth: {\it Cooling and
Low Energy State Preparation for 3-local Hamiltonians are
FQMA-complete}, LANL-preprint quant-ph/0305050.

\bibitem{PSPACE} P. Wocjan, D. Janzing, Th. Decker, Th. Beth: {\it
Measuring 4-local n-qubit observables could probabilistically solve
PSPACE}, LANL-preprint quant-ph/0308011.

\end{thebibliography}
\end{document}